\documentclass[
 reprint,nofootinbib,
 amsmath,amssymb,
 aps,
]{revtex4-2}

\usepackage{graphicx}
\usepackage{dcolumn}
\usepackage{bm}
\usepackage{xcolor}
\usepackage[colorinlistoftodos]{todonotes}

\usepackage{soul}

\begin{document}

\preprint{APS/123-QED}

\title{Control of autoresonant plasma beat-wave wakefield excitation}

\author{M.~Luo${}^1$, C.~Riconda${}^2$,  I.~Pusztai${}^1$, A.~Grassi${}^2$, J.~S.~Wurtele${}^3$, T.~F\"ul\"op${}^1$}
 
\affiliation{%
 ${}^1$ Department of Physics, Chalmers University of Technology,  G\"{o}teborg, SE-41296, Sweden
}%
\affiliation{
${}^2$ LULI, Sorbonne Université, CNRS, École Polytechnique, CEA, 75252 Paris, France
}%
\affiliation{
 ${}^3$ Department of Physics, University of California, Berkeley, California 94720, USA
}%

\begin{abstract}

Autoresonant phase-locking of the plasma wakefield to the beat frequency of two driving lasers offers advantages over conventional wakefield acceleration methods, since it requires less demanding laser parameters and is robust to variations in the target plasma density. Here, we investigate the kinetic and nonlinear processes that come into play during autoresonant plasma beat-wave acceleration of electrons, their impact on the field amplitude of the accelerating structure, and on acceleration efficiency. Particle-in-Cell simulations show that the process depends on the plasma density in a non-trivial way but can be reliably modeled under specific conditions. Beside recovering previous fluid results in the deeply underdense plasma limit, we demonstrate that robust field excitation can be achieved within a fully kinetic self-consistent modeling. By adjusting the laser properties, we can amplify the electric field to the desired level, up to wave-breaking, and efficiently accelerate particles; we provide suggestions for optimized laser and plasma parameters. This versatile and efficient acceleration scheme, producing electrons from tens to hundreds of $\rm MeV$ energies, holds promise for a wide range of applications in research industry and medicine.    

\end{abstract}

\maketitle

%

\section{Introduction}
The plasma beat-wave accelerator (PBWA), first proposed by Tajima and Dawson \cite{Tajimaprl1979}, is based on driving relativistic plasma waves by the ponderomotive force of the beat wave of two -- typically several picoseconds long -- laser pulses. Recently there has been a renewed interest in the PBWA scheme as an alternative to the prevailing laser wakefield acceleration (LWFA) scheme driven by a single fs-pulse \cite{leemans2006gev,ke2021,oubrerie2022controlled,PhysRevLett.128.164801,zhu2023} since it allows the acceleration mechanism to be efficient on a wider range of plasma and laser parameters. For example, the PBWA allows relaxed constraints on laser diffraction \cite{PhysRevAccelBeams.26.061301}, favors trapping and electron acceleration by utilizing near-critical densities \cite{photonics9070476}, and can be combined with an appropriate plasma channel to modify the phase velocity of the plasma wave \cite{plasma6010003}.  

However, the PBWA scheme suffers from an intrinsic limitation on the plasma wave amplitude because of the detuning due to the nonlinear increase of the plasma wavelength, leading to  a maximum  in the achievable electric field amplitude, of $E_{\rm RL}=(16a_1a_2/3)^{1/3}E_0$ -- the Rosenbluth-Liu (RL) limit~\cite{Rosenbluthprl1972}. Here, $E_0=m_ec\omega_{\rm pe}/e$ is the cold, nonrelativistic wave-breaking field, with $c$ the speed of light in vacuum, and $a_{1}$ and $a_{2}$ the amplitudes of the two co-propagating laser beams in terms of the normalized vector potential $a = e A/m_ec$. 

The nonlinear detuning can be controlled by introducing a slowly varying parameter, such as a frequency chirp or a density gradient. An appropriate choice, above a threshold value, for this parameter brings the system into resonance and allows it to remain phase-locked over an extended period, even in the presence of variations in  system parameters. This phenomenon, known as \emph{autoresonance} \cite{Fajansajp2001}, can be harnessed to overcome the RL limit of wakefield excitation \cite{DMG, lindbergprl, lindbergpop}. Autoresonance has been  studied and exploited in a variety of nonlinear systems, such as nonlinear stimulated Raman scattering \cite{Chapmanprl, Yaakobi2008, Luopop}, the nonlinear excitation of electron plasma waves \cite{friedland2020jpp, Munirovpre}, and ion-acoustic waves \cite{stefan1991,Friedlandpre2014, Munirovprr}.  

While the LWFA scheme has shown great promise in providing high acceleration gradients on the order of $\rm GV/cm$ and achieving $\rm GeV$-level energy gain \cite{leemans2006gev,ke2021,zhu2023}, it typically requires laser intensities on the order of $10^{19}\rm W/cm^2$ (normalized laser amplitude $a\geq 2$) and pulse compression to the order of the plasma wavelength (from tens to hundreds of $\rm fs$) to achieve optimal acceleration \cite{luwei2007,wenz2020physics}. However, in many industrial and medical applications \cite{photonics9070476}, sub-$\rm GeV$ or $\rm GeV$ energy levels are unnecessary. The autoresonant PBWA scheme appears as a potential alternative that can reach high accelerating fields at significantly lower laser intensities (around $10^{17}\rm W/cm^2$), and with reduced requirements on pulse compression. The scheme does not rely on focusing optical elements such as axicons \cite{PhysRevAccelBeams.26.061301}, and can operate at higher densities than the scheme proposed in \cite{Jakobsson2021} allowing higher acceleration gradients at lower normalized laser intensities. As we shall demonstrate in this article, by adjusting the laser duration or chirp rate, precise control over the electric field amplitude and electron beam energy can be achieved. The robustness and stability of the autoresonant PBWA scheme makes it also well suited in multi-staging configurations, coupled to other plasma-laser acceleration schemes (see e.g.~\cite{PhysRevResearch.5.013115} and references therein). Besides tolerating sizable uncertainties in density and laser frequency the scheme does not rely on the application of focusing optical elements, such as an axicon. 
In addition to electron acceleration applications, the long trains of periodic large-amplitude density oscillations achievable through  autoresonant PBWA could serve as a controllable moving-grating for manipulating light \cite{lehmann,caterina,wu}. The PBWA scheme has also been proposed as a platform for accelerating heavier particles, such as muons \cite{Peano, Peano_2008, Peano_2009}. 

To establish the realistic utility of the PBWA scheme in the applications mentioned, it is crucial to go beyond a simplified fluid description, and to explore its parametric dependences. In this paper, we use fully kinetic Particle-In-Cell (PIC) simulations to revisit autoresonance in the PBWA scheme. Taking previous studies based on a cold electron fluid model \cite{lindbergprl, lindbergpop} as a starting point, we explore a regime where both fluid nonlinearities and kinetic effects are important. Going well beyond the early kinetic study of PBWA by~\cite{Ghizzo}, we perform a systematic study  of parametric dependences on plasma density, chirp rate, laser intensity, and initial frequency shift in the lasers, and provide insights to further physics content of the process, such as fluid nonlinearities. We show that the frequency chirp allows an  efficient control of the wave amplitude as well as the self-injection of the particles.  
We also delineate under which circumstances and with respect to which quantities the cold electron fluid modeling can provide guidance. Apart from being reasonably accurate in the range of sufficiently low densities for which the self-consistent evolution of laser-plasma system can be neglected, the fluid model can also provide guidelines on the optimal chirp conditions even when fluid nonlinearities and kinetic effects are present.  In all regimes, the efficiency of this process is found to be insensitive to deviations of the density from its optimal value, thereby providing a robust wakefield excitation scheme under a broad range of conditions.

Our paper is organised as follows: in Sec.~\ref{model} the equation describing the plasma wave excitation under the PBWA scheme is numerically solved in the fluid and quasi-static approximation. The value of the optimal chirp parameter needed to overcome the RL limit is discussed, and the dependence of the autoresonant process on  the laser intensity and chirp rates, as well as variations in plasma density and other laser parameters, are discussed.  In Sec.~\ref{kinetic_simulation}, these results are then compared with fully nonlinear and kinetic simulations of the autoresonant PBWA process performed with the PIC code {\sc Smilei} \cite{DEROUILLAT2018351}, and the self-injection of the background electrons is discussed in Sec.~\ref{injection}. Finally, the conclusion and discussion are given in Sec.~\ref{conclusion}.


%

\section{Fluid modeling \label{model}}

We start by briefly revisiting the one-dimensional (1D) fluid modeling of the autoresonant wakefield excitation phenomenon, and we provide representative numerical results that will be compared to kinetic results in Sec.~\ref{kinetic_simulation}. Here we assume that the plasma -- consisting of a cold, collisionless, relativistic electron fluid, and a neutralizing, immobile ion background --  is strongly underdense, so that the phase velocity $v_p$ and group velocity $v_g$ of the plasma wave are close to the speed of light $c$, and  $\gamma_p^2=(1-\beta_p^2)^{-1}\approx (\omega_0/\omega_{\rm pe})^2\gg1$, where $\beta_p=v_p/c$. In the PBWA scheme, the wakefield is driven by the ponderomotive force of the beating wave of two co-propagating laser beams $\bar{a}_1=a_1\cos(\phi_1)$ and $\bar{a}_2=a_2\cos(\phi_2)$, where $\phi_l=k_lx-\omega_l t$, with $l=1,2$. 
The frequencies of two co-propagating lasers ($\omega_1$ and $\omega_2$) are detuned from each other by approximately the electron plasma frequency, i.e., $\omega_1-\omega_2\approx \omega_{\rm pe}\equiv(n_ee^2/\epsilon_0 m_e)^{1/2}$, where $n_e$ is the electron density, $m_e$ and $-e$ are the electron mass and charge, respectively, and $\epsilon_0$ is the vacuum permittivity. 

To induce the autoresonant growth of the wakefield \cite{lindbergprl}, a linear frequency chirp can be introduced in $\bar{a}_1$, i.e., $\omega_1=\omega_0+\alpha(t-t_0)$, where $\omega_0$ is the reference or central frequency, $\alpha$ is the chirp rate, and $t_0$ sets the initial frequency shift, $-\alpha t_0$. The frequency of the second laser beam is shifted by the electron plasma frequency, i.e., resulting in an initial frequency difference given by $\Delta\omega(0)=\omega_{\rm pe}-\alpha t_0$. By applying a negative chirp rate, the resonance can be swept slowly around the time  $t=t_0$. The ponderomotive force can be written as $a^2=\bar{a}^2+\epsilon\cos\psi(\xi)$, with $\bar{a}^2=(a_1^2+a_2^2)/2$, $\epsilon=a_1a_2$ is the normalized beat amplitude, and $\psi(\xi)=\int_0^{\xi}[\omega_1(\xi^{\prime})-\omega_2(\xi^{\prime})]d\xi^{\prime}$, with $\xi=\omega_{\rm pe}(t-x/v_g)$  the co-moving position.  This allows us to formulate  the quasi-static evolution of the normalized electrostatic potential $\phi=e\Phi_e/m_ec^2$ in the following way~\cite{lindbergpop},
\begin{equation}
\frac{\partial^2\phi}{\partial\xi^2}=\frac{1}{2}\left[\frac{1+\bar{a}^2+\epsilon\cos\psi(\xi)}{(1+\phi)^2}-1\right]. 
\label{wakefield_equation1}
\end{equation}
Note that this dimensionless equation does not have any explicit electron density dependence. From its solution, the longitudinal electric field can be calculated as $E_L =( \omega_{\rm pe} m_e c/e) (\partial/\partial\xi)\phi$. 

It is important to emphasize that Eq.~(\ref{wakefield_equation1}) does not include kinetic effects, such as wave-breaking, electron energization, or plasma heating, nor the self-consistent evolution of the plasma and the laser properties. Such effects can significantly impact the wakefield excitation, especially for the relatively long laser pulses employed in the PBWA scheme~\cite{forslund1985prl}. In the following, we   solve Eq.~(\ref{wakefield_equation1}) numerically to obtain an overview of the autoresonant PBWA and to provide the basis for  comparisons with PIC simulation results in Sec.~\ref{kinetic_simulation}. 

We consider a laser pulse of duration $T_{\rm dura}$ equal to $80\pi/\omega_{\rm pe}$, unless stated otherwise. The ponderomotive force profile is chosen so that it remains nearly constant during the wakefield excitation process, and is taken as 8$^{\rm th}$ super-Gaussian profile. We set $t_0=22.5\pi/\omega_{\rm pe}$, which allows sweeping the beat frequency through resonance. This choice of $t_0$ ensures that the resonance occurs when the laser intensity has already reached the plateau of the super-Gaussian profile. The laser duration can be chosen to avoid long-pulse effects that break the coherence of the wakefield.  For example, for the parameters considered here, the two-stream instability \cite{Mora1988prl} is expected to break the coherence of the wakefield over typical time scales of the inverse ion plasma frequency $\sim 1/\omega_{\rm pi}$, with $\omega_{\rm pi}= \omega_{\rm pe}(ZM_i/m_e)^{1/2} $, with $M_i$ the ion mass. For a singly ionized He plasma and, with  $T_{\rm dura} \lesssim 5/\omega_{\rm pi}$, this corresponds to $T_{\rm dura}\lesssim 100\pi/\omega_{\rm pe}$.

%

\subsection{Effects of the ponderomotive force and chirp rate}

\begin{figure}[htbp]
    \centering
    \includegraphics[width=0.9\linewidth]{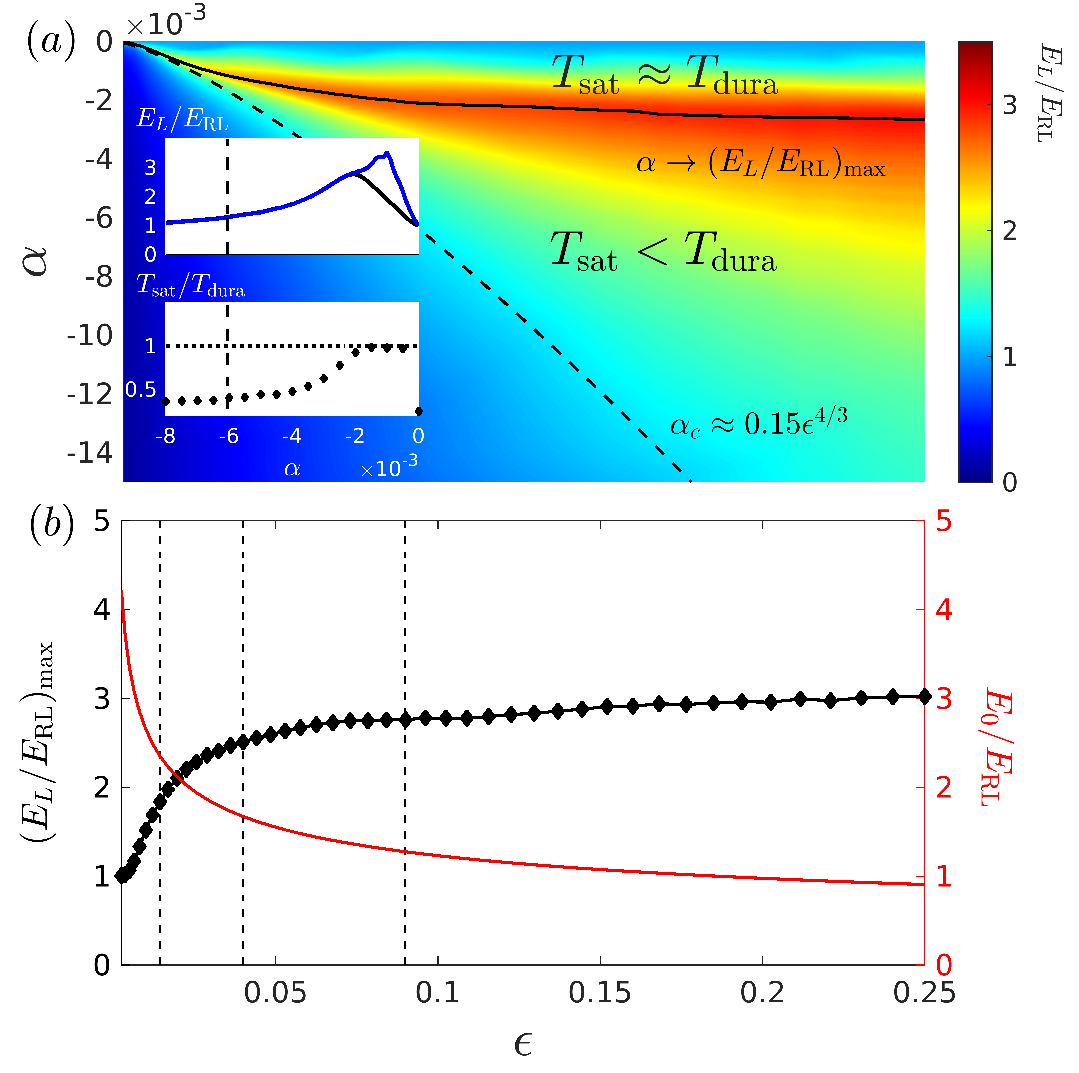}
    \caption{(a) Saturated electric field $E_L$ normalised to the RL field as a function of $\alpha$ and  $\epsilon=a_1a_2$ obtained from the numerical solution of Eq.~(\ref{wakefield_equation1}). Here  $T_{\rm dura}\omega_{\rm pe}=80\pi$. The black dashed line represents the critical chirp rate $\alpha_c\approx-0.15\epsilon^{4/3}$, and the black solid line indicates the optimal chirp rate choice to achieve the highest electric field for a given $\epsilon$. The top panel in the inserted figure displays $E_L$ as a function of chirp rate for $\epsilon=0.09$. The black (blue) line corresponds to  $T_{\rm dura}\omega_{\rm pe}=80\pi$ ($160\pi$). In the bottom panel of the insert the ratio between the saturation time $T_{\rm sat}$ of the autoresonant PBWA process and the laser duration  is shown as function of $\alpha$ for $T_{\rm dura}\omega_{\rm pe}=80\pi$. (b) The black diamond line represents the highest electric field that can be achieved by autoresonant PBWA as function of $\epsilon$, and the red line indicates the cold, nonrelativistic, wave-breaking value $E_0$ (normalized to $E_{\rm RL}$) as function of $\epsilon$.}  
\label{scan-intensity} 
\end{figure}

By numerically solving Eq.~(\ref{wakefield_equation1}) with different ponderomotive strength $\epsilon$ and chirp rates $\alpha$, Fig.~\ref{scan-intensity}(a) shows the saturated value of the longitudinal electric field carried by the wakefield normalized to the RL field,  $E_L/E_{\rm RL}$.
The solid black line corresponds to the chirp rate that yields the maximum electric field for a given ponderomotive force strength, denoted as $\alpha \rightarrow (E_L/E_{\rm RL})_{\rm max}$, while the dashed black line indicates the critical value $\alpha_c=-0.15\epsilon^{4/3}$ \cite{lindbergpop}, above which the absolute value of the chirp is too large to be effective in driving the autoresonance.

We illustrate the dependence of $E_L/E_{\rm RL}$ on the laser duration in the inset of Fig.~\ref{scan-intensity}(a), where the value of the normalized field versus $\alpha$ is shown for a given value of the ponderomotive force, here $\epsilon = 0.09$, for two different values of the laser duration: $T_{\rm dura}=80\pi/\omega_{\rm pe}$ (black curve), that is the same value used for Fig.~\ref{scan-intensity}(a), and $T_{\rm dura}=160\pi/\omega_{\rm pe}$ (blue curve). As a reference, the vertical dashed line indicates the critical value $\alpha_c$. 
The electric field increases with $\alpha$ and then decreases for higher chirp rates after having reached a maximum. 
As we can see, the two curves overlap, except for small absolute values of the chirp, in which cases increasing the laser duration results in higher values of the field. To understand the impact of the laser duration, in the lower panel of the inset, we plot the saturation time of the wakefield excitation normalized to the laser duration, with $T_{\text{dura}}=80\pi/\omega_{\rm pe}$, obtained from the solution of Eq.~(\ref{wakefield_equation1}). 
For a given laser duration, a longer saturation time leads to a higher value of the electric field (black line in the top panel of the inset) up to $T_{\rm sat}\simeq T_{\text{dura}}$ and, subsequently, the electric field starts decreasing. 
The optimal value of $\alpha$ thus corresponds to a saturation time equal to $T_{\text{dura}}$. 

In the fluid model, raising $T_{\rm dura}$ further leads to an increase in the achievable value of $E_L$ for the small $|\alpha|$ values as discussed in Ref.~\cite{lindbergpop}. However constraints on the maximum laser duration will be given in practice by physical processes not included in the model Ref.~\cite{lindbergpop}, thus justifying our initial choice of the laser duration.   

In Fig.~\ref{scan-intensity}(b), we report the maximum electric field $(E_L/E_{\rm RL})_{\rm max}$ (black diamonds) obtained for a given ponderomotive force strength $\epsilon$ obtained by using the optimal chirp (analogous to the  solid black line in Fig.~\ref{scan-intensity}(a)). The autoresonant excitation reaches values of the  plasma wave field significantly larger than the RL limit, increasing steadily up to $\epsilon\lesssim 0.03$ after which the ratio $(E_L/E_{\rm RL})_{\rm max}$ stays at roughly $3$ for the chosen time duration. 
As $\epsilon$ increases kinetic effects become more important, up to the point where the wave-breaking of the plasma wave limits its amplitude. 
Thus in the same Fig.~\ref{scan-intensity}(b) we show the ratio of the cold wave-breaking field value over the RL-limit as a function of the ponderomotive force, given by $E_0/E_{\rm RL}=(16\epsilon/3)^{-1/3}$.
As $\epsilon$ increases, the ratio between $E_0$ and $E_{\rm RL}$ decreases, and for $\epsilon > 0.02$, $(E_L)_{\rm max}$ becomes larger than $E_0$. In principle, this should define the limit of validity of the fluid theory, which is applicable only for values of $\epsilon$ such that $E_0/(E_L)_{\rm max}\geq1$. 
However, as we  show in the following section were we consider full kinetic simulations for three different values of $\epsilon=0.0144$, $0.04$, and $0.09$ (reported in Fig.~\ref{scan-intensity}(b) as vertical dashed line for reference), the fluid theory  provides useful estimates and insights even at large values of $\epsilon$.

%

\subsection{The robustness of the autoresonant PBWA}

\begin{figure}[htbp]
    \centering
    \includegraphics[width=0.9\linewidth]{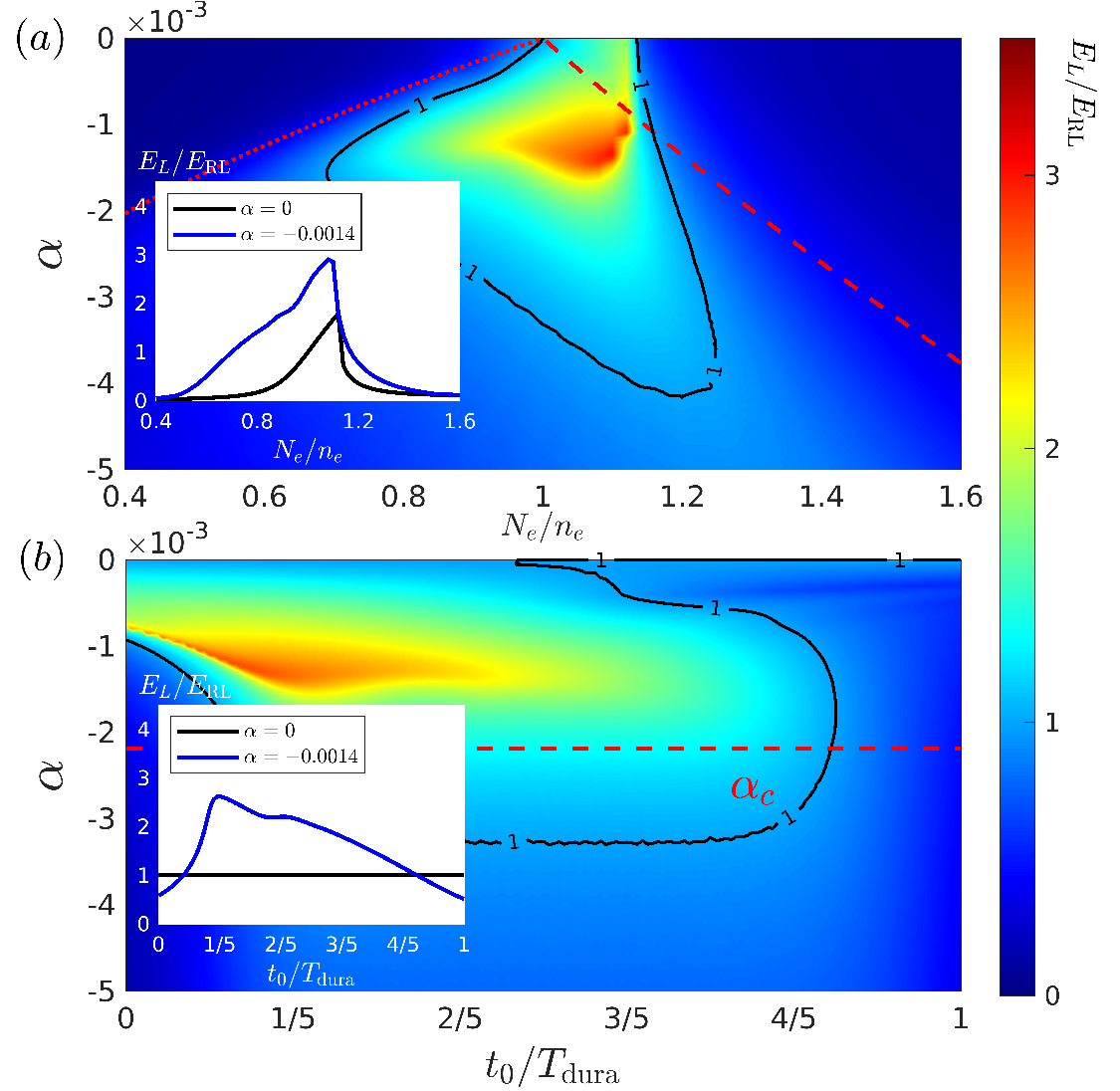}
    \caption{(a) Saturated electric field $E_L$ as a function of chirp rate $\alpha$ and density variation $N_e/n_e$, obtained from the numerical solution of Eq.~(\ref{wakefield_equation1}) with the ponderomotive force $\epsilon=a_1a_2=0.04$. Here, $n_e$ represents the ideal matching density, while $N_e$ represents the actual density. The red dashed and dotted lines represent $N_e/n_e=\Delta\omega(0)^2$ and $N_e/n_e=\Delta\omega(T)^2$, respectively. The inserted figure illustrates the saturated electric field $E_L$ as a function of density variation $N_e/n_e$ with chirp rate $\alpha=0$ (black line) and $\alpha=-0.0014$ (blue line). (b) Saturated electric field $E_L$ as a function of the chirp rate $\alpha$ and the time-instant $t_0$ when the matching condition is met. The horizontal dashed red line represents the critical chirp rate $\alpha_c$. The inserted figure displays the saturated electric field $E_L$ as a function of $t_0$ with chirp rate $\alpha=0$ (black line) and $\alpha=-0.0014$ (blue line).}  
\label{scan_den_chirp} 
\end{figure}

Next we investigate the robustness of the PBWA process concerning uncertainties or variations in the value of the plasma density, still in the framework of the fluid approach. The laser frequencies are carefully selected to satisfy a matching condition for the nominal electron density $n_e$, i.e., $\omega_1-\omega_2=\omega_{\rm pe}(n_e)$. However, the actual plasma density $N_e$ may deviate from this nominal value. As a result, the right-hand side of Eq.~(\ref{wakefield_equation1}) is multiplied by $N_e/n_e$. 
In Fig.~\ref{scan_den_chirp}(a), we present the maximum electric field as a function of the chirp rate and $N_e/n_e$, assuming $a_1=a_2=0.2$, or $\epsilon=0.04$, while keeping the initial frequency difference $\Delta\omega(0)=\omega_{\rm pe}-\alpha t_0$, with other parameters related to the ponderomotive force held fixed.
The autoresonant PBWA demonstrates a remarkable enhancement of the electric field over a wide range of the $\alpha$--$N_e/n_e$ parameter space, as illustrated in Fig.~\ref{scan_den_chirp}(a), where $E_L/E_{\rm RL}=1$ is highlighted by the black contour. Particularly, for chirp rates $|\alpha|\in[0.0005,0.0018]$ and plasma density ratios $N_e/n_e\in[0.85,1.15]$, the saturated amplitude surpasses twice the RL limit. This indicates a significant robustness of the wakefield excitation process. In contrast, for zero chirp rate, appreciable excitation of the plasma wave is only observed in a narrow vicinity of $N_e/n_e=1$. 
For the sake of illustration, in the inset in Fig.~\ref{scan_den_chirp}(a), we show lineouts of the underlying figure at $\alpha=0$ (black curve) and $\alpha=-0.0014$ (blue curve). The case with negative chirp shows a significant enhancement of the field over a wider range of the plasma densities.

For $N_e/n_e>\Delta\omega(0)^2$ or $N_e/n_e<\Delta\omega(T_{\rm dura})^2$, the matching condition is never satisfied throughout the entire duration of the laser interaction. 
Consequently, the plasma wave is not efficiently driven. The regions above the red dashed and dotted lines in Fig.~\ref{scan_den_chirp}(a) correspond to these situations, respectively. 
Notice that there exists a finite region above the red dashed curve where the electric field amplitudes exceed unity. This is achieved through a two-step process: First, there is a near-resonant excitation of a plasma wave with a small but finite amplitude, driven by the finite bandwidth around the exact resonance. When the amplitude is sufficiently close to the exact resonance, the corresponding nonlinear upward shift of the plasma wavelength brings the system into resonance. From this point on, the autoresonant process takes over, to drive sizable wave amplitudes. This phenomenon explains why the effect is only observed for $N_e/n_e>1$, as the nonlinear upward shift of the initially lower wavelength can bring the system into resonance. 

In Fig.~\ref{scan_den_chirp}(b), the impact of $t_0$ is shown, in conditions in which the matching condition is met, for $a_1=a_2=0.2$ and $N_e/n_e=1$. In the ideal case in which the ponderomotive force duration $T_{\rm dura}$ is arbitrarily long, the specific value of $t_0$ would not have a significant impact on the wakefield excitation process (assuming no instabilities to limit the useful time-range). However, in reality, the time needed to satisfy the matching condition influences the evolution history and, consequently, the saturation amplitude of the wakefield.
We observe that the dependence of $E_L/E_{\rm RL}$ on $t_0$ is relatively weak over a wide range of $t_0$ values. For example, the inset in Fig.~\ref{scan_den_chirp}(b) provides an illustration of the robustness of the wakefield excitation process with respect to the variation of $t_0$ for $\alpha=0$ (black line) $\alpha=-0.0014$ (blue line). Only at the two ends of the reported $t_0$ range we observe a notable decrease in the saturated amplitudes. This indicates the presence of an optimal range of $t_0$ values, specifically $t_0/T_{\rm dura} \in [1/10,2/5]$, for which the wakefield excitation reaches its highest amplitudes. This optimal range corresponds to chirp rates $|\alpha|\in[0.0005,0.0018]$, which in turn corresponds to initial frequency differences $\Delta\omega(0)\in [1.013\omega_{\rm pe},1.181\omega_{\rm pe}]$, for the chosen ponderomotive force strength. 
This confirms the robustness of the proposed mechanism.

%

\section{Kinetic study of the autoresonant PBWA \label{kinetic_simulation}}

To investigate the autoresonant PBWA process using kinetic simulations, we perform 1D simulations with {\sc Smilei} PIC code. The simulation setup is detailed in Appendix~\ref{appendix_setup}. The physical parameters are identical to the ones used in the fluid simulations. 
%
%
%
\begin{figure}[htbp]
    \centering
	\includegraphics[width=1.\linewidth]{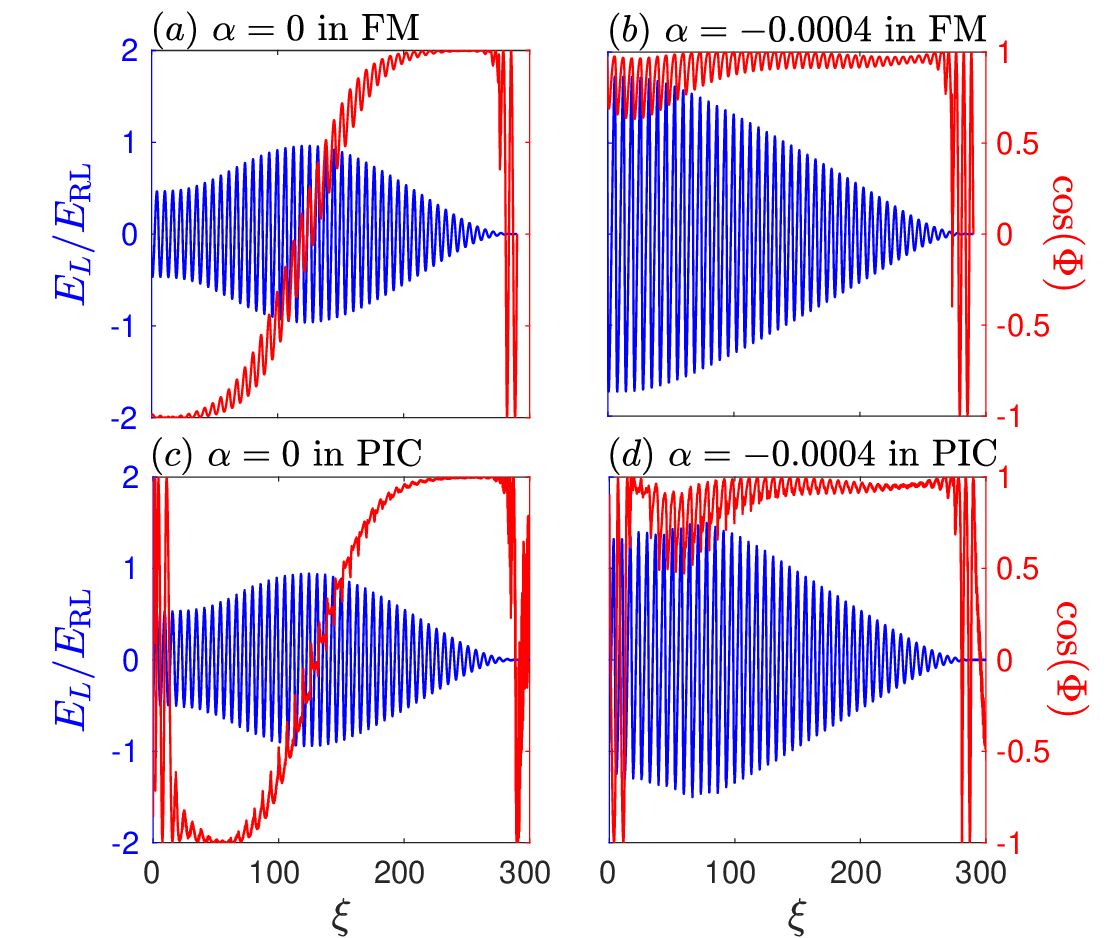}
    \caption{The top panels display the numerical solutions of Eq.~(\ref{wakefield_equation1}) for laser intensities $a_1=a_2=0.12$ with chirp rates $\alpha=0$ in panel (a) and $\alpha=-0.0004$ in panel (b). The spatial profile of the electric field $E_L$ is represented by the blue line, while the cosine value of the relevant phase $\Phi$ is shown by the red line, where $\Phi(\xi)=\psi(\xi)-\phi_L(\xi)$, with $\phi_L={\rm Arg}(E_L)$, and $\psi(\xi)$ is the phase difference between the two laser beams. The bottom panels show the same quantities obtained at $\omega_{\rm pe}t\approx600$ by the PIC simulations, with chirp rate $\alpha=0$ in panel (c) and $\alpha=-0.0004$ in panel (d).} 
\label{normalization-chirp} 
\end{figure}
We initially select $a_1=a_2=0.12$, where the maximum electric field $E_L$ is expected to be larger than the RL limit but smaller than the wave-breaking threshold $E_0$, as indicated in Fig.~\ref{scan-intensity}(b) by the leftmost dashed line. The background plasma density is chosen to be $n_e/n_{cr}=0.0004$ and the total simulation time is $T_{\text{sim}}\omega_{\text{pe}} = 240\pi$.
Figure~\ref{normalization-chirp} illustrates the wakefield profile (blue lines) and phase evolution (red lines) obtained either by solving Eq.~(\ref{wakefield_equation1}) (top panels) or as a result of the kinetic simulations (bottom panels). In the fluid model, the phase is defined as $\Phi(\xi)=\psi(\xi)-{\rm Arg}(E_L(\xi))$, whereas in the kinetic simulation, it is defined as $\Phi(\xi)=\varphi_1(\xi)-\varphi_2(\xi)-{\rm Arg}(E_L(\xi))$, with the phase of laser beam $1 (2)$ denoted by $\varphi_{1(2)}$. Figs.~\ref{normalization-chirp}(a) and (c) represent the cases without a chirp, while Figs.~\ref{normalization-chirp}(b) and (d) depict the case with a chirp rate of $\alpha=-0.0004$. Good agreement is obtained between the fluid model results and kinetic simulation results. In both cases, the wakefield grows when the phase $\Phi$ is locked around 0 or 2$\pi$, corresponding to $\cos(\Phi)=1$. 
The highest electric field, denoted as $E_L$, reaches a value that is around 1.5 times the RL limit, with the help of autoresonance. 

\begin{figure}[htbp]
    \centering
	\includegraphics[width=1\linewidth]{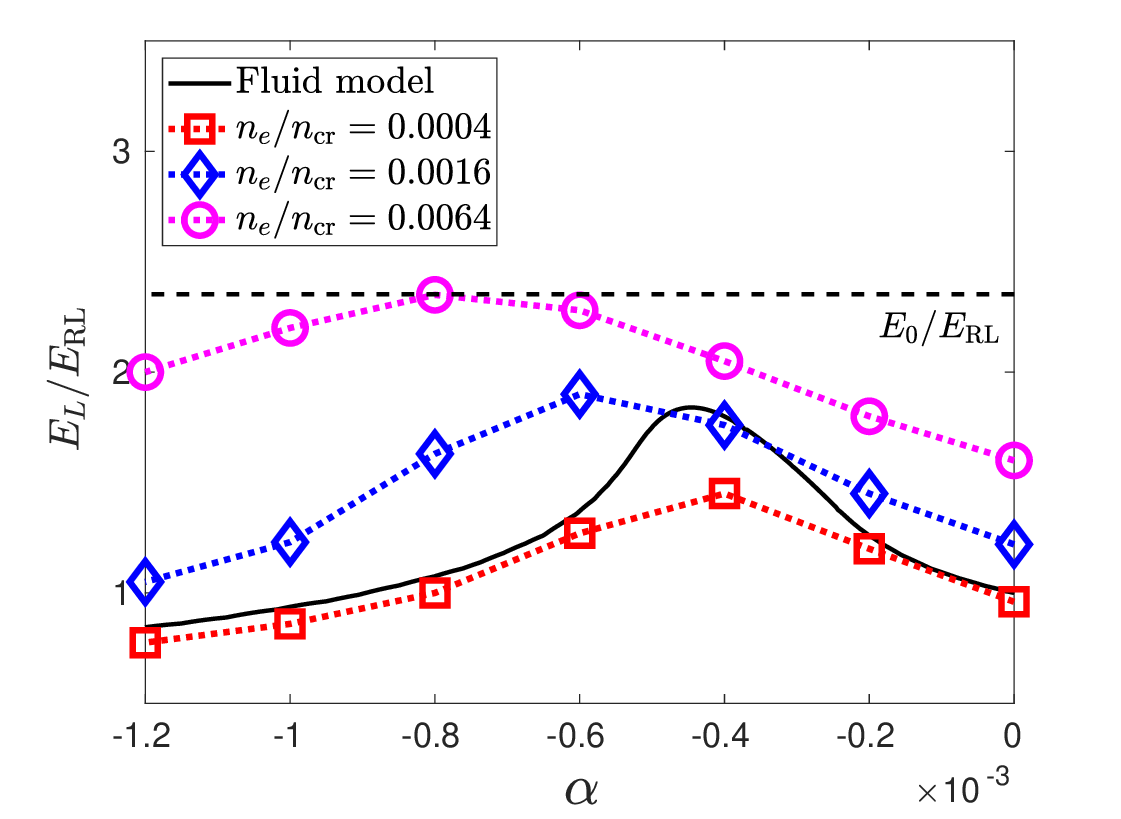}
    \caption{The maximum amplitude of the plasma wave $E_L$ as a function of the chirp rate $\alpha$, for $a_1=a_2=0.12$. The black line represents the numerical solution of Eq.~(\ref{wakefield_equation1}), while the colored symbols represent the PIC simulation results at different plasma densities.}  
\label{beat0.12_chirp} 
\end{figure}

\begin{figure*}[htbp]
    \centering
	\includegraphics[width=1\linewidth]{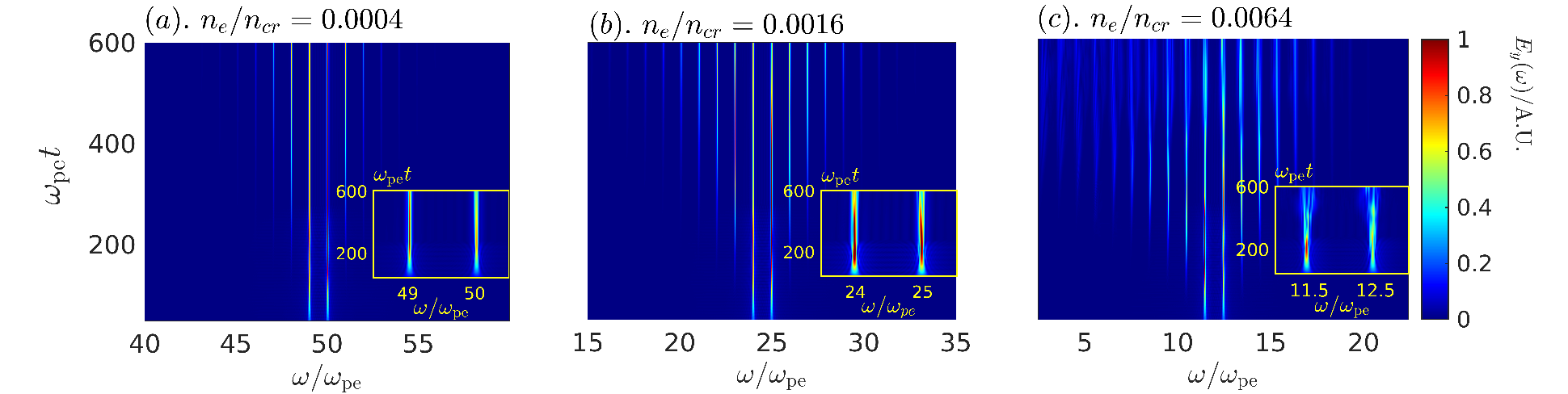}
    \caption{The frequency spectrum of the electromagnetic fields evolving in time is depicted for three distinct plasma densities. The laser intensities are fixed at $a_1=a_2=0.12$, and no chirp is applied to the laser beams. The three plasma density cases considered are $n_e/n_{cr}=0.0004$ in (a), $n_e/n_{cr}=0.0016$ in (b), and $n_e/n_{cr}=0.0064$ in (c). The inserted figures offer detailed views of the frequency evolution of the two primary laser beams, $a_1$ and $a_2$, over time during the simulation.}  
\label{laser_scattering} 
\end{figure*}

\begin{figure*}[htbp]
    \centering
	\includegraphics[width=1\linewidth]{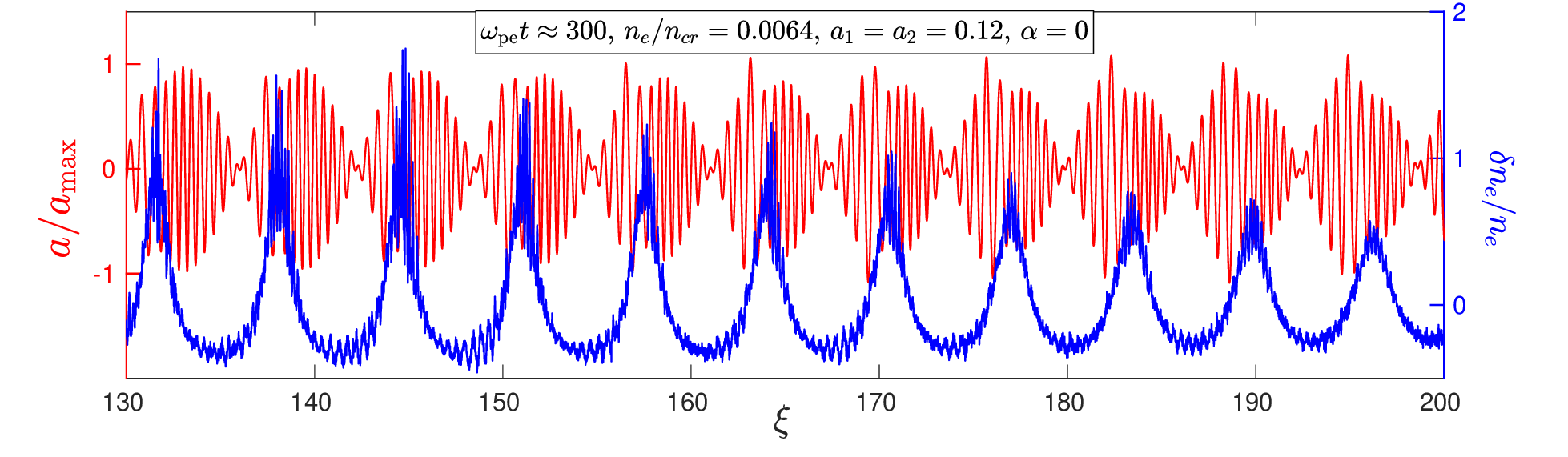}
    \caption{The spatial profiles of the electromagnetic field (represented by the red line) and the relative density perturbation $\delta n_e/n_e$ (represented by the blue line) for laser intensities $a_1=a_2=0.12$, with no chirp applied to the laser beams. The plasma density is $n_e/n_{cr}=0.0064$. The plot shows the time instance $\omega_{\rm pe}t\approx 300$, when the scattering instability is relatively weak.}  
\label{density_perturbation} 
\end{figure*}

The maximum of the electric field $E_L$ obtained from the PIC simulations, normalized to the RL limit $E_{\rm RL}$ is shown (red squares) in Fig.~\ref{beat0.12_chirp} as a function of the chirp rate for a density $n_e/n_{cr}=0.0004$ and normalized laser amplitude $a_1=a_2=0.12$. The resulting curve closely follows the solid black line obtained by the fluid model, the only difference being that the peak value of the PIC simulations is lower than the corresponding value of the fluid model. The physics behind this lower value will be discussed later in this section.

The fluid model (i.e.~Eq.~(\ref{wakefield_equation1})) has no explicit dependence on the plasma density, but $E_{\rm RL}$ increases with the square root of the density, so that the impact of the density on the absolute value of the peak fields is only included in the normalization. This is not the case for the PIC simulations. 
We consider two additional densities, i.e.,~$n_e/n_{cr}=0.0016$ and $n_e/n_{cr}=0.0064$, for which  
the kinetic results of $E_L/E_{\rm RL}$ are presented by blue diamonds and pink circles in Fig.~\ref{beat0.12_chirp}, respectively. The dependence of $E_L/E_{\rm RL}$  on the chirp within the kinetic approach is qualitatively similar to the fluid one, but its value is systematically larger (or approximately equal in one case) than the value obtained by the fluid model for the same chirp. The peak of $E_L/E_{\rm RL}$ is obtained for larger absolute values of the chirp parameter as the density increases. 
However, as discussed in more detail in the following of this section, at high density, the system enters a nonlinear regime in which the structure of the plasma wave field becomes more irregular; thus, an increase in field amplitude by itself is not the only figure of merit for a stable acceleration structure. 
For lower plasma densities, e.g., $n_e/n_{cr}=0.0001$, the qualitative behaviour is similar to the case $n_e/n_{cr}=0.0004$.


In the kinetic simulations, the plasma and the laser fields are evolved self-consistently, which can induce a modification of the laser beating and of the ponderomotive driver, depending on the plasma density. Indeed, the difference between the fluid and kinetic simulations can be attributed to several underlying factors:  the appearance of Stokes and anti-Stokes scattering of the two laser beams, as discussed in Ref.~\cite{walton2002large,deng,Coverdale}; the modification of laser frequencies due to local density perturbations; and, when present, particle trapping.  
 
In Fig.~\ref{laser_scattering}, spectrograms of the electromagnetic waves are presented, for cases in which no chirp is applied to the laser beams\footnote{In the Appendix~\ref{appendixA}, we also show the frequency spectrum of the electromagnetic fields, with a chirp $\alpha=-0.0004$ and   normalized laser amplitudes $a_1=a_2=0.12$.}. The figures correspond to different plasma densities, ranging from $n_e/n_{cr}=0.0004$ in Fig.~\ref{laser_scattering}(a) to $n_e/n_{cr}=0.0064$ in Fig.~\ref{laser_scattering}(c). Note that in our kinetic simulations, the temporal and spatial dimensions are normalized by the plasma frequency and plasma wavenumber, and we consider a fixed simulation time of $T_{\text{sim}}\omega_{\text{pe}} = 240\pi$.  When considering different plasma densities this results in different interaction times when normalised  to  the laser frequency $T_{\text{sim}}\omega_0 = 12000\pi$ for $n_e/n_{\text{cr}} = 0.0004$, $T_{\text{sim}}\omega_0 = 6000\pi$ for $n_e/n_{\text{cr}} = 0.0016$, and $T_{\text{sim}}\omega_0 = 3000\pi$ for $n_e/n_{\text{cr}} = 0.0064$. 

In the frequency spectra, clear Stokes/anti-Stokes side-bands with a regular interval ($\Delta\omega\approx\omega_{\rm pe}$) are observed. As the plasma density increases, 
these side-bands become more pronounced. The appearance of the side-bands in the frequency spectrum indicates the presence of a scattering phenomenon due to the interaction between the laser beams and the plasma. The growth of the Stokes/anti-Stokes scattering depends on the plasma density, e.g., $\gamma_{\rm s}/\omega_{\rm pe}\propto \sqrt{n_e}$ \cite{kruer2019physics, walton2002large, deng, Coverdale, Michel2023} (more details are given in Appendix~\ref{appendixB}). 

The insets in Fig.~\ref{laser_scattering} zoom in on the evolution of the frequencies of the two laser beams 
for the same time interval as the background images. As the plasma density is increased, the main spectral lines of the two laser beams broaden. To explain the broadening (visible in the inset of Fig.~\ref{laser_scattering}(c)), the spatial profile of the vector potential is shown in Fig.~\ref{density_perturbation}. The figure plots the beating of the two laser beams resulting in the ponderomotive drive (red curve) and  the relative density perturbation $\delta n_e/n_e$ (blue curve). At the time corresponding to the data in Fig.~\ref{density_perturbation}, $\omega_{\rm pe}t\approx 300$, the Stokes and anti-Stokes components are still weak. 

As shown in Fig.~\ref{density_perturbation}, the plasma wave is characterized by a strong density perturbation. This can lead to up/down-shifts in the laser frequencies, as also discussed in Ref.~\cite{RevModPhys.81.1229}. Specifically, the negative gradient component of the density perturbation up-shifts the laser frequency, while the positive gradient component down-shifts the frequency, following the relationship $\delta\omega\propto-d(\delta n_e/n_e)/d\xi$. This can be seen in Fig.~\ref{density_perturbation}: the oscillations of the electromagnetic field are elongated in the region where the density perturbation exhibits a positive gradient and shortened when the opposite gradient is present. This effect becomes particularly pronounced starting at  $\xi\approx150$--$170$ and going toward the left. 

All these effects are responsible for the difference in the maximum field amplitude for $\alpha=0$ observed in Fig.~\ref{beat0.12_chirp}. At larger absolute values of $\alpha$,  the electric field $E_L$ becomes stronger so that all these nonlinearities become even more pronounced. 
A simple correction to the linear plasma wave wavelength $\lambda_p$ as $\lambda_{np} = \lambda_{p}(1+3(E_L/E_0)^2/16)~$ \cite{1990a,1990b,RevModPhys.81.1229}
allows us to estimate the nonlinear wavenumber spectral evolution of the plasma wave. More details can be found in Appendix~\ref{appendixC}, where we explicitly show the validity of this approximation by comparing the evolution of the numerically obtained wavenumber with the formula above. 

Note that, in all considered cases, except for the case at the lowest density and no chirp, even if the electric field $E_L$ is technically below the wave-breaking limit, large density perturbations are present, inducing self-injection of electrons. This induces a further nonlinearity that can affect the ponderomotive potential~\cite{PhysRevLett.28.417,Ghizzo}; it will be discussed in the next section. 

To summarize, we showed that the quasi-static approximation used in Eq.~(\ref{wakefield_equation1}) will become invalid when the density is large enough to produce significant scattering of the laser or very large density perturbations. However, as we will discuss in detail in the next section, in addition to the fluid theory being predictive if the plasma is sufficiently underdense, the stable and optimal electron acceleration scenarios correspond to regimes where these fluid nonlinearities are weak.

\begin{figure}[htbp]
    \centering
	\includegraphics[width=1\linewidth]{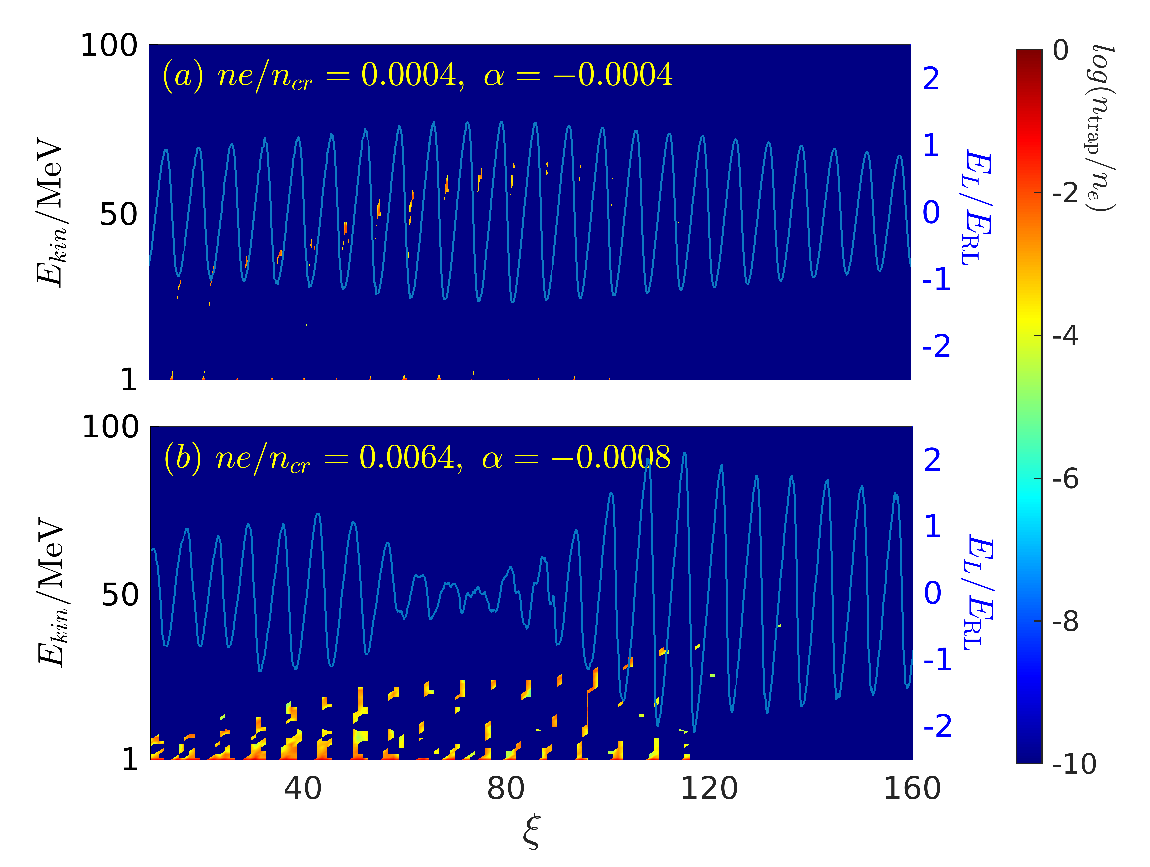}
    \caption{The kinetic energy spectrum of the particles and the electric field $E_L$ normalized by the RL limit $E_{\rm RL}$, taken at time $\omega_{\rm pe}t\approx450$ for $a_1=a_2=0.12$. (a) corresponds to $\alpha=-0.0004$ and a density of $n_e/n_{cr}=0.0004$;  (b) uses a chirp rate $\alpha=-0.0008$ and a density $n_e/n_{cr}=0.0064$.}  
\label{phase_space} 
\end{figure}

\begin{figure*}[htbp]
    \centering
	\includegraphics[width=1\linewidth]{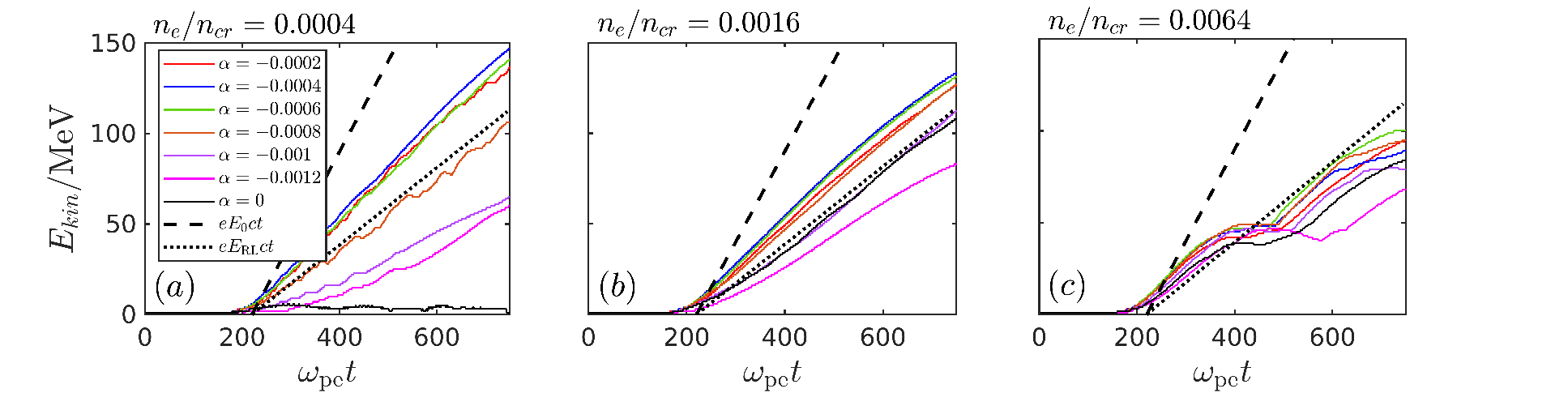}
    \caption{Time evolution of the kinetic energy of the particles reaching the highest energies, for different chirp values. The black dashed and dotted lines represent $E_{kin,0}=eE_{0}ct$ and $E_{kin,\rm RL}=eE_{\rm RL}ct$, respectively. They represent the kinetic energy obtained by the electrons accelerated by the wave-breaking value $E_0$ and the RL limit value $E_{\rm RL}$, respectively. The plasma density is chosen as $n_e/n_{cr}=0.0004$ in (a), $n_e/n_{cr}=0.0016$ in (b), and $n_e/n_{cr}=0.0064$ in (c).}  
\label{maximum_energy_time} 
\end{figure*}

\begin{figure}[htbp]
    \centering
	\includegraphics[width=1\linewidth]{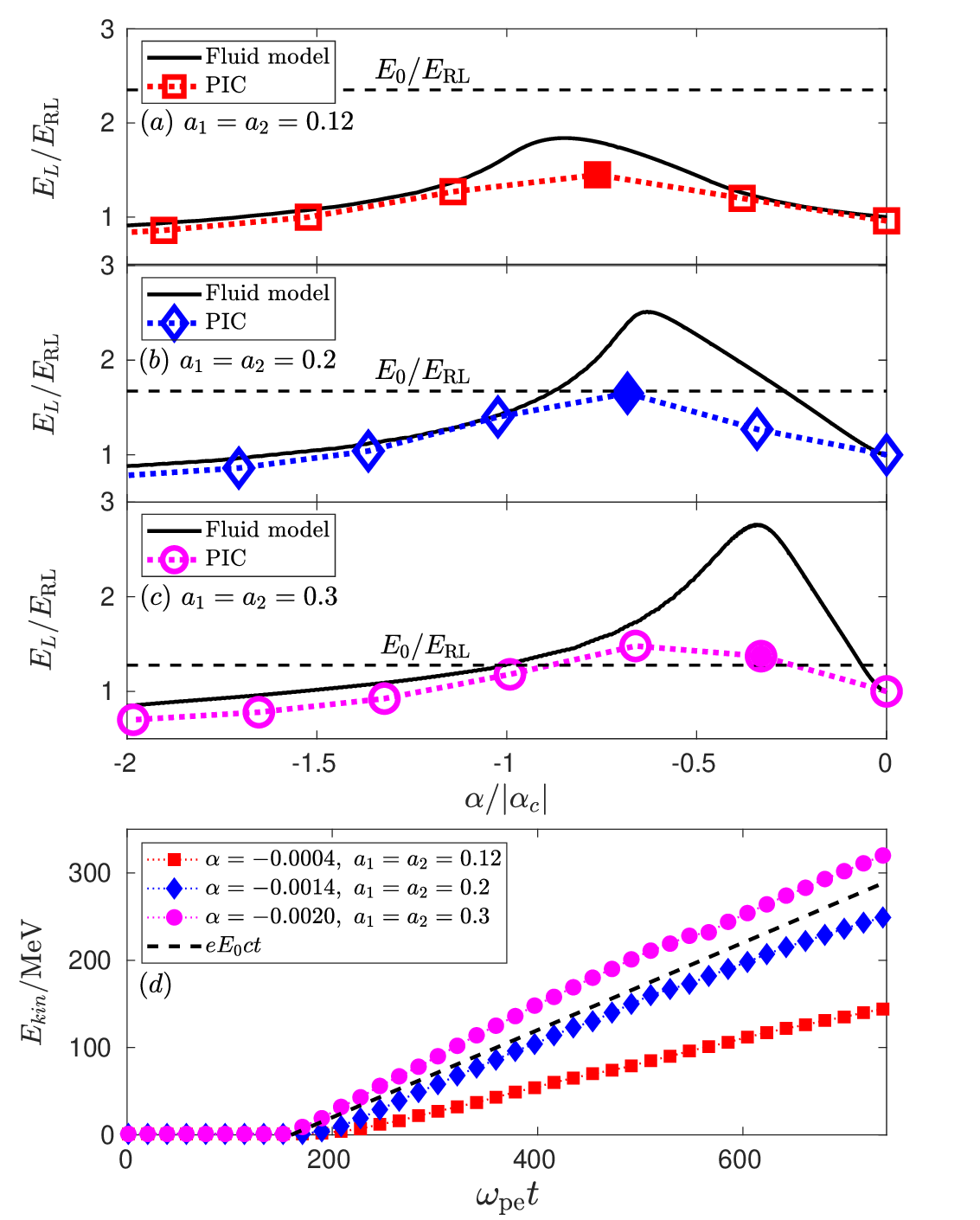}
    \caption{The maximum amplitude of the plasma wave $E_L$ is shown as a function of chirp rate,  for the plasma density $n_e/n_{cr}=0.0004$. The laser intensities are $a_1=a_2=0.12$ in (a), $a_1=a_2=0.2$ in (b), and $a_1=a_2=0.3$ in (c). The black solid lines represent the numerical solution of Eq.~(\ref{wakefield_equation1}), while the colored dotted lines represent the PIC simulation results. The horizontal black dashed lines show the wave-breaking value. In (d), the evolution of the kinetic energy of the most energetic electrons is shown for the cases exhibiting the most efficient acceleration at various laser intensities. For $a_1=a_2=0.12$, the most efficient case corresponds to $\alpha =$ -0.0004. For $a_1=a_2=0.2$, it corresponds to $\alpha =$ -0.0014. Lastly, for $a_1=a_2=0.3$, it corresponds to $\alpha =$ -0.002. The black dashed line corresponds to the kinetic energy obtained for the wave-breaking value $E_0$.}
\label{camparison_fluid_kinetic} 
\end{figure}

%

\section{Self-injection in the autoresonant PBWA \label{injection}}

In Sec.~\ref{kinetic_simulation}, we analyzed the electric field $E_L$ carried by the plasma wave driven within the autoresonant PBWA scheme through PIC simulations. The results are qualitatively consistent with the fluid model, especially when the plasma density remains sufficiently underdense. However, at higher plasma densities, nonlinearities associated with the laser evolution appear. This creates large, localized density peaks that can efficiently trap particles. Because of that, on the one hand, the plasma wave structure becomes inhomogeneous and, on the other hand, beam loading by particles contributes to damping the wave. This means that despite the possibility of transiently achieve amplitudes that exceed those predicted by the fluid model, the highest density cases are not necessarily optimal for acceleration over long distances. Establishing a reliable accelerator hinges not only on achieving a high electric gradient but also on maintaining a stable accelerating field structure. 

To investigate the injection and acceleration strength of these structures, in Fig.~\ref{phase_space}, we depict the kinetic energy spectrum of the electrons along with $E_L/E_{\rm RL}$, as function of $\xi$ at $\omega_{\rm pe}t\approx450$ for $a_1=a_2=0.12$. In Fig.~\ref{phase_space}(a) the chirp rate is $\alpha=-0.0004$ and the plasma density $n_e/n_{cr}=0.0004$. In this case, the plasma wave maintains a coherent and regular structure, achieving an electric field $E_L$ of around $1.5E_{\rm RL}$ with the assistance of autoresonance. It is noteworthy that this coherent structure endures over an extended period. Particles are self-injected and experience acceleration through the electric gradient, attaining energies of  $\sim60$ MeV. 
In Fig.~\ref{phase_space}(b), the chirp rate is $\alpha=-0.0008$ and the plasma density is $n_e/n_{cr}=0.0064$,  corresponding to the cases with the highest field value in Fig.~\ref{beat0.12_chirp}. 
As we can see, although at this time the transient normalized field ($E_L/E_{\rm RL}\approx2$) is larger than in the previous case, the plasma wave is not homogeneous, and the high field value is sustained for a shorter period of time. 
Although more particles can be accelerated due to the higher background plasma density, the highest particle energy is about half of that in Fig.~\ref{phase_space}(a), e.g., $\sim$30 MeV.
As discussed in the previous section, the plasma wave amplitude and the wavenumber spectra evolution with time exhibit a strong dependence on the density due to the various nonlinearities in the system. The coherence of the plasma wave is reduced at higher plasma density (see Appendix~\ref{appendixC} for more details).

Once particles are self-injected and phase locked in the longitudinal electric field, we can assume that the highest kinetic energy is approximately given by $E_{\text{kin}}=eE_{L}v_pt$, where $v_pt$ represents the acceleration length. In the present case of a very underdense plasma, $v_p$ can be approximated by the speed of light $c$. When self-injection occurs, the electric field can approach, or even reach, the wave-breaking threshold, leading to an upper limit in the kinetic energy. This can be expressed as $E_{kin,0}=eE_{0}ct=m_ec^2\omega_{\text{pe}}t=\omega_{\text{pe}}t/2\text{(MeV)}$. 
In the kinetic simulations, time is normalized to the plasma frequency, hence, the kinetic energy $E_{\text{kin}}$ as a function of $\omega_{\text{pe}}t$ is the same for different plasma densities.
For reference, we also consider the highest kinetic energy that would be obtained if the acceleration was limited by the RL limit electric field, $E_{kin,\rm RL}=eE_{\rm RL}ct=\omega_{\text{pe}}t/2(16\epsilon/3)^{1/3}\text{(MeV)}$. These two limits can be considered to assess the stability and strength of the accelerator structure.

The kinetic energy of the electrons reaching the highest energy is shown in Fig.~\ref{maximum_energy_time}, for all cases shown in Fig.~\ref{beat0.12_chirp}. The black dashed and dotted lines represent $E_{kin,0}=eE_{0}ct$ and $E_{kin,\rm RL}=eE_{\rm RL}ct$, respectively. Note that the cases shown in Fig.~\ref{phase_space}(a) and Fig.~\ref{phase_space}(b) correspond to the  blue line in Fig.~\ref{maximum_energy_time}(a) and the orange line in Fig.~\ref{maximum_energy_time}(c), respectively.

For a plasma density $n_e/n_{\text{cr}}=0.0004$ in Fig.~\ref{maximum_energy_time}(a), if no chirp is applied, the electric field is limited to $E_{\rm RL}$ and cannot accelerate particles (solid black line, $\alpha=0$). Owing to the increase of the field due to autoresonant excitation, particle acceleration becomes significant, especially for values of chirp for which the electric field experiences a large enhancement. The accelerating strength for different chirp rates is consistent with the behavior of the electric field shown in Fig.~\ref{beat0.12_chirp}. 

In Fig.~\ref{maximum_energy_time}(b), where the plasma density is increased to $n_e/n_{\text{cr}}=0.0016$, the peak accelerated particle energy  is still  consistent  with the variation of the maximum field as a 
function of the chirp parameter seen in Fig.~\ref{beat0.12_chirp}. However, due to the fluid nonlinearities previously discussed, the maximum energy obtained in Fig.~\ref{maximum_energy_time}(b) is lower than the one obtained in Fig.~\ref{maximum_energy_time}(a). We further see that at  $\omega_{\text{pe}}t\approx600$, the accelerating strength starts to decline. Finally, as shown  in Fig.~\ref{maximum_energy_time}(c), the acceleration stops at early times due to the loss of coherence of the electric field structure; this  is visible, for instance, in Fig.~\ref{phase_space}(b).
In conclusion,  as  the plasma density is increased (from Fig.~\ref{maximum_energy_time}(a) to Fig.~\ref{maximum_energy_time}(c)), the overall acceleration stability and strength decreases. The theoretical upper limit on the acceleration length is the \emph{dephasing length}, $L_d$, over which a relativistic electron outruns the accelerating phase of the laser field by a quarter wavelength; it is given by $L_d k_p=\pi\omega_0^2/\omega_{\rm pe}^2$ \cite{luwei2007,wenz2020physics}. A lower plasma density gives a longer dephasing length, as seen in Table~\ref{dephasing}, showing the normalized dephasing lengths for the densities used in Fig.~\ref{maximum_energy_time}. 
The $L_d k_p$ values are directly comparable to the $\omega_{\rm pe}t$ scale of Fig.~\ref{maximum_energy_time}. In Fig.~\ref{maximum_energy_time}(c) we see that the acceleration efficiency drops significantly earlier than the dephasing length (that would be above $\omega_{\rm pe}t=600$, since the acceleration starts around $\omega_{\rm pe}t=200$), and it is caused by losing the coherence of the accelerating structure. Thus wave coherence appears as a more stringent limiting factor than dephasing.

\begin{table}[htbp]
\begin{center}
\begin{tabular}{c|c|c|c}
    $n_e/n_{cr}$ &  $L_dk_p$&$E_{\rm the}$ [GeV]&$L_d$ [mm]\\
    \hline
    0.0004 &7853& 4&50\\
    0.0016&1963&0.98&6.25\\
    0.0064&490&0.25&0.78
\end{tabular}
\end{center}
\caption{Normalized dephasing length $L_dk_p$, theoretically  attainable maximum energy $E_{\rm the}=m_ec^2\pi\omega_0^2/\omega_{\rm pe}^2$, and absolute dephasing length for a $800\,\rm nm$ wavelength laser, for different values of the plasma density.\label{dephasing}}
\end{table}

In order to identify the optimal parameters of laser intensity and chirp rate required to obtain a stable field close to the wave-breaking value, we fix the plasma density at $n_e/n_{cr}=0.0004$ and vary the intensity. The value of $E_L/E_{\rm RL}$ as a function of the chirp rate normalized to the absolute value of the critical chirp rate, i.e., $\alpha/|\alpha_{c}|$, is reported in Fig.~\ref{camparison_fluid_kinetic} for $a_1=a_2=0.12$ in (a), $a_1=a_2=0.2$ in (b), and $a_1=a_2=0.3$ in (c). The critical chirp rate $\alpha_c=-0.15\epsilon^{4/3}$ represents the limit above which (in absolute value) the chirp becomes ineffective in driving autoresonace. The black solid lines in Fig.~\ref{camparison_fluid_kinetic} represent the numerical solutions obtained from Eq.~(\ref{wakefield_equation1}), while the colored symbols correspond to the  PIC simulation results. For reference, the wave-breaking field is indicated by a horizontal black dashed line.

We find that, for a given density, the electric field $E_L$ obtained in the PIC simulations can be approximately predicted by the numerical solutions of Eq.~(\ref{wakefield_equation1}) as long as the latter is below wave-breaking. Whenever the fluid prediction for the field exceeds the wave-breaking limit, the autoresonant excitation drives the plasma wave close to that limit, as shown in Fig.~\ref{camparison_fluid_kinetic}(b-c).

In Fig.~\ref{camparison_fluid_kinetic}(d), we show the energy evolution of the most energetic electrons for the chirp values
that the electron can obtain the highest energy in Fig.~\ref{camparison_fluid_kinetic}(a-c)(i.e., the cases indicated by solid symbols). In this figure, the black dashed line corresponds to the kinetic energy obtained using the wave-breaking value $E_0$. Although the highest absolute energy is obtained at the highest intensity, the intermediate case with $a_1=a_2=0.2$ is the one in which the autoresonance achieves the largest ratio of the accelerating field to its RL limit as shown in Fig.~\ref{camparison_fluid_kinetic}(b-c).

For optimal plasma and laser parameters, the maximum attainable energy is limited by the dephasing length and the theoretically highest accelerating electric field (the wave-breaking field). Neglecting pump-depletion this energy is given by $E_{the}=m_ec^2\pi\omega_0^2/\omega_{\rm pe}^2$. We provide values of $E_{the}$ for a few normalized densties in Table~\ref{dephasing}. For instance, $E_{the}=4\,\rm GeV$ for $n_e/n_{cr}=0.0004$. However, in Fig.~\ref{camparison_fluid_kinetic}(d), the plotted time range is much shorter than that corresponding to the dephasing length ($k_pL_d=7853$).
It is expected that physics effects emerging in a higher dimensional treatment will pose a more stringent limit on the effective acceleration times compared to dephasing; this is left for future investigation.

%

\section{Conclusions \label{conclusion}}

We investigated the autoresonant Plasma Beating Wakefield Acceleration (PBWA) scheme by employing a chirped laser. First, the wakefield excitation equation is numerically solved under the fluid and quasi-static approximation, in order to discuss representative cases of the autoresonant behavior of a plasma wave, using the formalism and typical range of parameters of Ref.~\onlinecite{lindbergprl}. We show the relationship between the saturation time of the autoresonant excitation and the laser duration time. Since the duration of the acceleration -- and so the meaningful laser pulse length -- in a real system is limited by physical processes not captured by the fluid model, we find that  for relatively weak chirp strengths, the saturation time of the process and the optimal chirp rate will depend on the laser duration. A systematic scan of the chirp rates allows us to identify the highest electric fields achieved in PBWA, using the fluid model. This acceleration scheme appears very robust with respect to variations in plasma density and frequency shift between the two co-propagating lasers. 

To establish the practical utility of the scheme, and to gain additional physics insights, we go beyond the fluid formalism and conducted a series of one-dimensional PIC simulations.  Unlike the fluid model that is quasi-static and uses imposed laser fields,  the kinetic simulations account for the self-consistent evolution of the plasma density and of the laser pulses propagating in the plasma, as well as kinetic effects. Specifically, as the plasma density is increased, we observe that the Stokes or anti-Stokes scattering of the laser beams is enhanced, and the up-shift or down-shift of the laser frequencies due to density perturbation becomes more significant, further affecting the plasma wave dynamics. Despite this, for sufficiently underdense plasmas, such as $n_e/n_{cr}=0.0004$, these nonlinearities are weak, and the kinetic simulation results agree well with the fluid model results. For this density value, we found that the maximum enhancement of the electric field above the Rosenbluth-Liu (RL) limit is obtained for a normalized laser intensity of $a_1=a_2\approx 0.2$ with an appropriate chirp rate. In this case, the electric field is amplified exactly to the level of the cold, nonrelativistic wave-breaking electric field. Electrons are efficiently self-injected and accelerated. 

To make a connection to experiments using a Ti:sapphire CPA $800\,\rm nm$ laser, we note that the optimized parameters for the autoresonant PBWA scheme that we found in this work correspond to laser intensities of $I_1\approx I_2=8.5\times 10^{16}\ \rm W/cm^2$ and a plasma density of $n_e\approx 7\times 10^{17}\ \rm cm^{-3}$. The laser duration should be around $\approx 4$--$5\,\rm ps$, and the total bandwidth in the laser beams corresponds to $0.5$--$0.7\%$ due to the chirp. With these parameters, the electric field gradient can reach up to $\sim$ 80 GV/m (wave-breaking electric gradient), almost twice the Rosenbluth-Liu (RL) limit  gradient ($\sim$ 47 GV/m). The maximum kinetic energy can reach 256 MeV over an acceleration length of $L\approx3.5$ mm. Assuming a cylindrical geometry and the laser focal spot of $w_0\approx40 \rm \mu m$, and integrating the energy interval of $10\%$ of the highest kinetic energy, i.e., $\Delta E\approx25$ MeV, these accelerated high-energy particles will be concentrated in a region of transverse and longitudinal extension of $\sim15\mu m$, this scheme can yield a cumulative charge of approximately $30\,\rm pC$ (the corresponding density of the trapped particles is $n_{\rm trap}/n_e=$ 0.0403). In addition, considering the same laser parameters, the system is shown to be robust to variations in plasma density. Operating in a density range of $5\times 10^{17}\ \rm cm^{-3}$ to $8\times 10^{17}\ \rm cm^{-3}$, corresponding to $\pm 23\%$ variation in the density, the autoresonant PBWA scheme can still efficiently excite an electric field gradient above the RL limit; a supporting discussion is provided in Appendix~\ref{appendixD}.

It is essential to highlight that the optimal chirp rate can vary depending on the laser intensity. 
For instance, when considering a laser intensity of approximately $I_1 \approx I_2 = 3 \times 10^{16}\ \rm W/cm^2$, the required bandwidth in the laser beams due to the chirp is about $0.2\%$. However, in this case, the longitudinal electric field may not reach the level necessary for wave-breaking, although the density of the high-energy particles could be higher $\sim n_{\rm trap}/n_e=$ 0.073. If the laser intensity is increased to $I_1 \approx I_2 = 2 \times 10^{17}\ \rm W/cm^2$, a broader bandwidth of approximately $\sim 1\%$ in the laser beam is needed. Although this setup can drive a longitudinal electric field to the level of wave-breaking, it does  
not produce a higher accelerated charge when considering an acceleration length similar to the above mentioned optimal case (the corresponding density is $n_{\rm trap}/n_e=$ 0.046). 
Hence, the laser intensity should be carefully evaluated to optimize the trade-off between accelerated energy  and charge. 
Furthermore, strategies to successfully mitigate nonlinearities at higher plasma densities may be possible, and appear as an interesting avenue for further investigation to obtain the higher electric gradient and charges.    

GeV energy levels reachable with LWFA are often unnecessary in many industrial and medical applications, and for many facilities the corresponding requirements on laser intensity and pulse compression are out of reach. Autoresonant PBWA, which operates robustly already at intensities of $10^{17}\rm W/cm^2$, thus represents an interesting alternative. It provides precise control over the electric field amplitude through adjusting the laser duration or chirp rate, allowing for fine-tuning of the particle energy. The long, high amplitude plasma density oscillation trains could also serve as a controllable moving-grating for manipulating light.

We note that a two-dimensional kinetic analysis of the autoresonant PBWA indicate that, while additional rich physics (such as the appearance of Weibel-type instabilities and transverse filamentation of the wakefield) shows up, our conclusions stand the test of higher dimensionality. The detailed analysis in higher dimensions is left for a future publication. 

\begin{acknowledgments}
The authors are grateful for   fruitful discussions with Stefan H$\rm\Ddot{u}$ller and Ning Wang. This project received funding from the Knut and Alice Wallenberg
Foundation (Grant no.~KAW 2020.0111).  The computations were enabled
by resources provided by the Swedish National
Infrastructure for Computing (SNIC), partially funded
by the Swedish Research Council through grant
agreement no.~2018-05973. We thank the Berkeley-France Fund for support of this research.
\end{acknowledgments}


\appendix

%

\section{The fundamental simulation setup in kinetic study \label{appendix_setup}}

In the kinetic study, the simulation setup follows our previous baseline case in Sec.~\ref{model}, where two co-propagating laser beams with parallel linear polarization and identical intensities are employed. To prevent ion instabilities, the duration of the laser pulses $T_{\rm dura}$ is chosen to be $80\pi/\omega_{\rm pe}$, and the longitudinal pulse shape is modeled as an 8th-order super-Gaussian. Here $t_0=22.5\pi/\omega_{\rm pe}$, and a homogeneous plasma density is assumed throughout the simulation box. The ions are set to be immobile, as they are not important for the dynamics on the time scale of the laser duration~\cite{Mora1988prl}, as we verified with complementary simulations with mobile ions (not shown here). In the 1D simulations, the spatial resolution is chosen to be $dx=0.008 k_p^{-1}$, and the time step is adjusted accordingly to $c dt=0.9dx$, to satisfy the Courant–Friedrichs–Lewy condition. We initialized $100$ macro-particles per cell per species. 

\section{The frequency spectrum of the laser beams with chirp \label{appendixA}}

\begin{figure*}[htbp]
    \centering
	\includegraphics[width=1\linewidth]{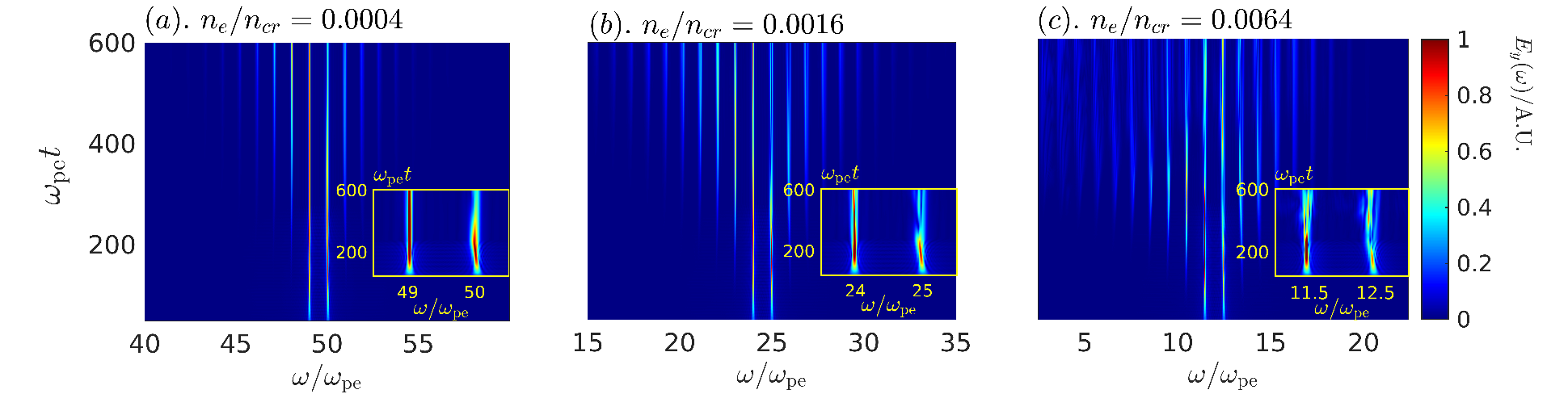}
    \caption{The frequency spectrum of the electromagnetic fields evolving in time is depicted for three distinct plasma densities. The laser intensities are fixed at $a_1=a_2=0.12$. The chirp is applied to the first laser beam with the chirp rate $\alpha=-0.0004$. The three plasma density cases considered are $n_e/n_{cr}=0.0004$ in (a), $n_e/n_{cr}=0.0016$ in (b), and $n_e/n_{cr}=0.0064$ in (c). The inserted figures offer detailed views of the frequency evolution of the two primary laser beams, $a_1$ and $a_2$, over time during the simulation.}  
\label{laser_scattering_chirp} 
\end{figure*}

In contrast to Fig.~\ref{laser_scattering}, where no chirp is applied, Fig.~\ref{laser_scattering_chirp} shows the frequency spectrum of the electromagnetic fields when a chirp is introduced. Specifically, a chirp rate of $\alpha=-0.0004$ is used, along with the laser intensities $a_1=a_2=0.12$. Notably, these three panels clearly show the presence of the frequency chirp carried by the first laser beam. After  $\omega_{\rm pe}t\approx300$,  the entire laser beams are located inside the moving window and different frequency components arising from the chirp are encompassed within the window. As a result, the spectrum of the first laser beam changes.  Nevertheless, the scattering instability of the two laser beams is still significant at higher plasma density, and, additionally, the large density perturbation resulting from the plasma wave shifts the frequencies of the two laser beams. This leads to the broader spectra shown in the inserted figure in Fig.~\ref{laser_scattering_chirp}.

%

\section{The first-order Stokes and anti-Stokes scattering of the laser beams \label{appendixB}}

\begin{figure}[htbp]
    \centering
	\includegraphics[width=0.95\linewidth]{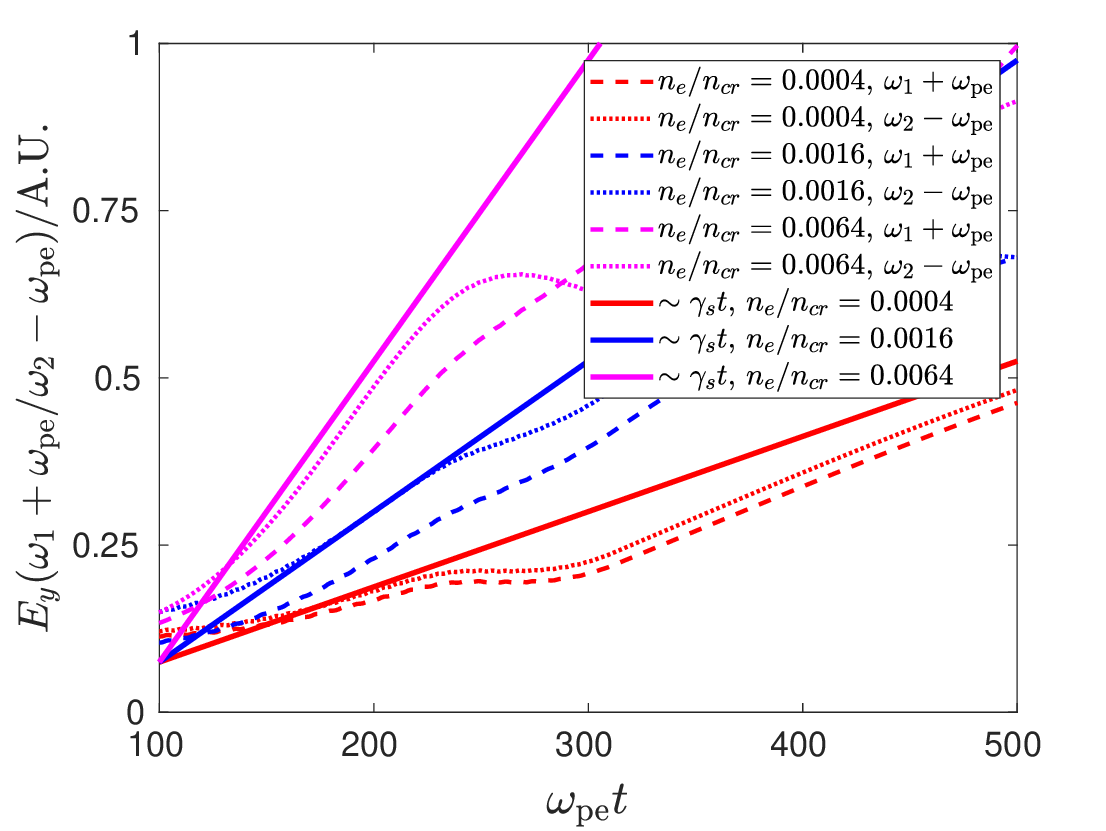}
    \caption{Integrating the scattered light amplitude over frequency ranges corresponding to first-order up-scattering and down-scattering of laser beams. Dashed lines indicate up-scattering, and dotted lines show down-scattering. Colors distinguish different plasma densities: red ($n_e/n_{\text{cr}} = 0.0004$), blue ($n_e/n_{\text{cr}} = 0.0016$), and pink ($n_e/n_{\text{cr}} = 0.0064$). The solid colored lines follow $\sim\gamma_st$, where the growth rates follow $\gamma_s/\omega_{\rm pe}(n_e/n_{cr}=0.0064)$=$2\gamma_s/\omega_{\rm pe}(n_e/n_{cr}=0.0016)$=$4\gamma_s/\omega_{\rm pe}(n_e/n_{cr}=0.0004)$.}  
\label{stokes_antistokes} 
\end{figure}

Figure~\ref{laser_scattering} shows the Stokes and anti-Stokes scattering of the laser beams under different plasma densities, where the frequency spectrum of the electromagnetic fields has a regular interval ($\Delta\omega\approx\omega_{\rm pe}$). Integrating the amplitude of the scattered light over the frequency spectrum $[\omega_1+\omega_{\rm pe}/2, \ \omega_1+3\omega_{\rm pe}/2]$ and $[\omega_2-3\omega_{\rm pe}/2, \ \omega_2-\omega_{\rm pe}/2]$, namely, the first-order up-scattering of the first laser beam and first-order down-scattering of the second laser beam, and giving the results shown in Fig.~\ref{stokes_antistokes}. The dashed colored lines indicate the first-order up-scattering, while the dotted colored lines represent the first-order down-scattering. Different colors are associated with different plasma densities, such as red for $n_e/n_{\text{cr}} = 0.0004$, blue for $n_e/n_{\text{cr}} = 0.0016$, and pink for $n_e/n_{\text{cr}} = 0.0064$. The dependence of the scattering on the plasma density is evident. The slope of these curves is well approximated by the solid colored lines, which represent $\sim\gamma_st$. The approximate growth rate $\gamma_s$ is calculated as follows. Following the calculation presented in Ref.~\cite{chapman2012autoresonance} and neglecting convection, we find  
\begin{equation}
    \gamma_s\sim\partial_t a_s = \frac{e}{4m_e}\frac{k_L}{\omega_s}E_L^{*}a_{1,2}.
    \label{scattering}
\end{equation}
Here, $a_s$ represents the normalized field of the scattered light and its resonant frequency $\omega_s\approx\omega_0$ for the first-order scattering. The electric field $E_L$ can be approximated by the RL-limit value, i.e., $E_L=E_0(16a_1a_2/3)^{1/3}$. Substituting this into Eq.~(\ref{scattering}) yields
\begin{equation}
    \frac{\gamma_s}{\omega_{\rm pe}}\sim\frac{\sqrt{n_e/n_{cr}}}{4}a_{1,2}(16a_1a_2/3)^{1/3}.
    \label{scattering-rate}
\end{equation}
The growth rates satisfy the relation 
\begin{eqnarray}
    \left. \frac{\gamma_s}{\omega_{\rm pe}} \right|_{n_e/n_{cr}=0.0064} = & \left. 2\displaystyle{\frac{\gamma_s}{\omega_{\rm pe}}} \right|_{n_e/n_{cr}=0.0016}
    \nonumber \\ = &  4\left.\displaystyle{\frac{\gamma_s}{\omega_{\rm pe}}} \right|_{n_e/n_{cr}=0.0004} \nonumber.
\end{eqnarray}

%

\section{The wavenumber spectrum of the plasma wave for different plasma densities and chirp rates \label{appendixC}}

\begin{figure}[htbp]
   \centering
	\includegraphics[width=0.95\linewidth]{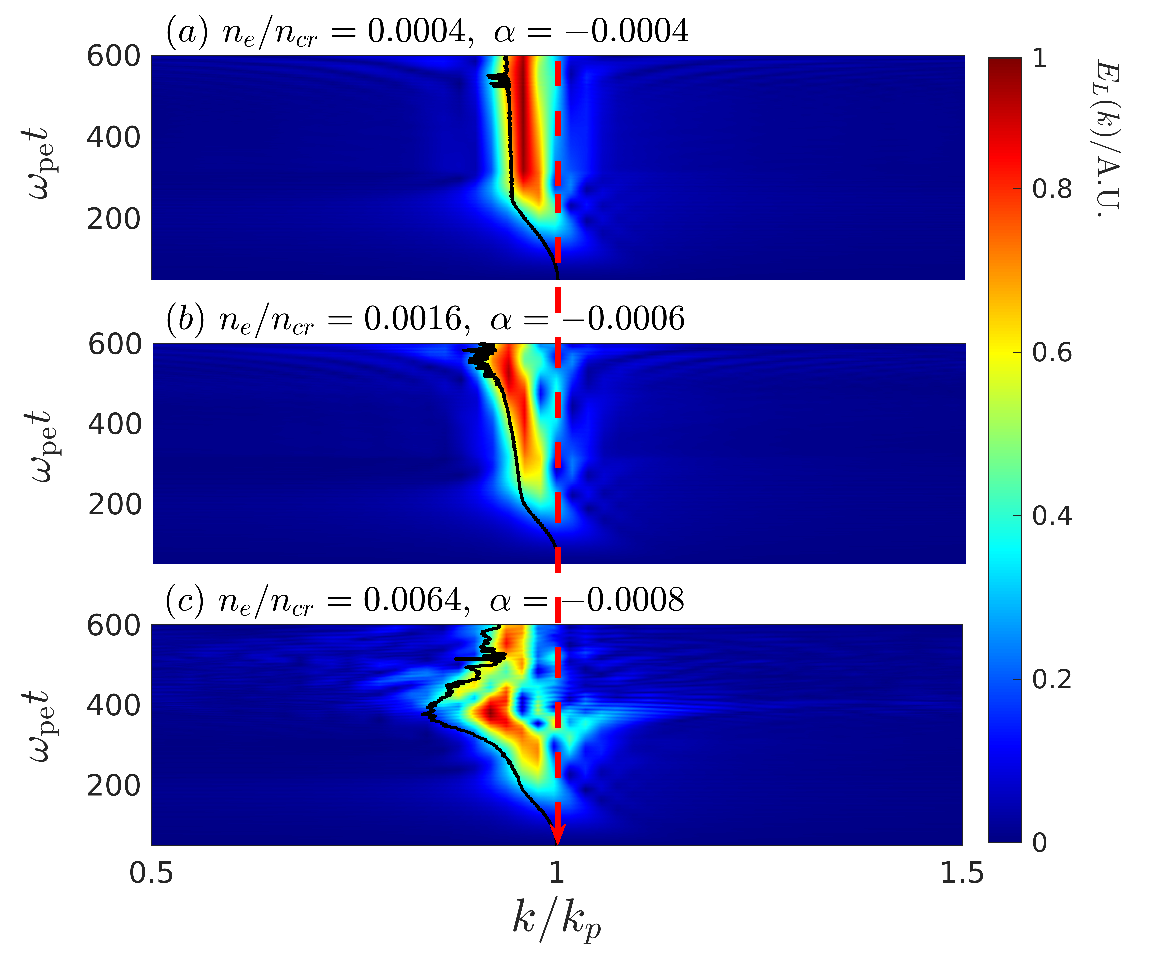}
    \caption{The wavenumber spectrum evolution of the plasma wave for laser intensities $a_1=a_2=0.12$ across different plasma densities and chirp rates. Specifically, for chirp rate $\alpha=-0.0004$ with plasma density $n_e/n_{cr}=0.0004$ in (a), $\alpha=-0.0006$ with $n_e/n_{cr}=0.0016$ in (b), and $\alpha=-0.0008$ with $n_e/n_{cr}=0.0064$ in (c). The solid black lines correspond to an analytical estimate of the maximum wavenumber shift based on the highest electric field $E_{L-\rm max}$ observed in the simulation.}  
\label{plasma-wave-coherence} 
\end{figure}

Figure~\ref{plasma-wave-coherence} illustrates the evolution of the wavenumber spectrum of the plasma wave for the cases corresponding to the peak values of the normalized electric field $E_L/E_{\rm RL}$, represented by the three color curves in Fig.~\ref{beat0.12_chirp}; that is, for the laser intensities $a_1=a_2=0.12$, chirp rate $\alpha=-0.0004$ with the plasma density $n_e/n_{cr}=0.0004$, $\alpha=-0.0006$ with $n_e/n_{cr}=0.0016$, and $\alpha=-0.0008$ with $n_e/n_{cr}=0.0064$. Generally, due to the nonlinear wavelength shift in the plasma wave, the resonant nonlinear wavenumber $k_{np}$ is smaller than the linear wavenumber $k_{p}$. Meanwhile, as the plasma density increases from Fig.~\ref{plasma-wave-coherence}(a) to Fig.~\ref{plasma-wave-coherence}(c), the coherence of the plasma wave deteriorates.

However, it is possible to estimate the maximum wavenumber shift. Building on the insights of Ref.~\cite{1990a,1990b,RevModPhys.81.1229}, we can derive the nonlinear plasma wavelength as $\lambda_{np}=\lambda_p(1+3(E_L/E_0)^2/16)$, where $\lambda_p$ is the linear plasma wavelength. Consequently, the nonlinear plasma wavenumber can be computed as $k_{np}=2\pi/\lambda_{np}$. Substituting $E_L$ with the highest electric field $E_{L-\rm max}$ and plotting $k_{np}$ using the solid black line in Fig.~\ref{plasma-wave-coherence} produces a reasonably good agreement.

%

\section{The robustness of the autoresonant PBWA scheme to laser frequency and plasma density variations \label{appendixD}}

\begin{figure}[htbp]
   \centering
	\includegraphics[width=1\linewidth]{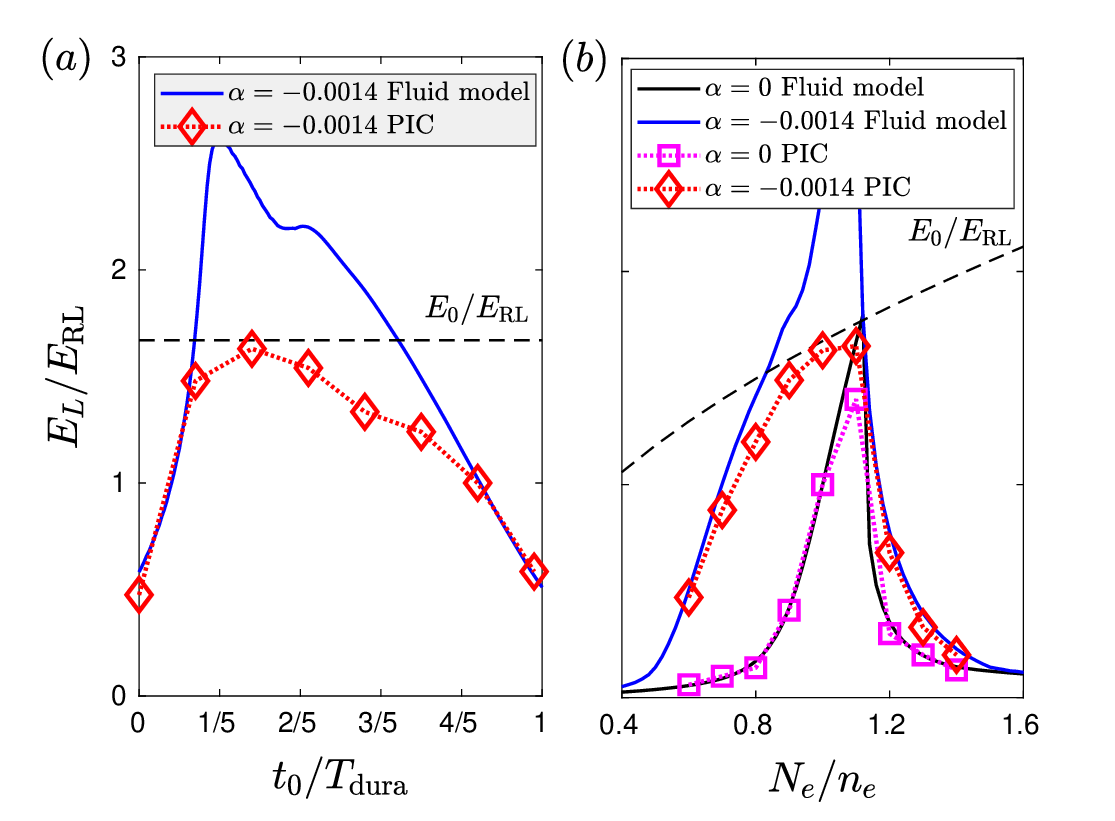}
    \caption{The dependence of the autoresonant PBWA laser on $t_0$ and $N_e/n_e$ are shown in panel (a) and (b), respectively, for intensities $a_1=a_2=0.2$. The red diamond (solid blue lines) represent the PIC simulation (fluid) results with chirp rate of $\alpha=-0.0014$. The black solid lines show the numerical solution of Eq.~(\ref{wakefield_equation1}) without chirp, and the black dashed line shows the wave-breaking value $E_0$ in the unit of RL limit value $E_{\rm RL}(N_e=n_e)$.}  
\label{robustness} 
\end{figure}

Fig.~\ref{robustness}(a) investigates the effect of varying $t_0$, which represents the specific moment when the resonance condition is met, while keeping the plasma density at the ideal matching density $n_e$. The laser intensities are $a_1=a_2=0.2$ and plasma density is $n_e/n_{cr}=0.0004$.  Fig.~\ref{robustness}(b) explores the impact of varying the plasma density $N_e$ relative to the ideal matching density $n_e$, while keeping time fixed at  $t_0/T_{\rm dura}\approx0.28$. In both panels, the solid black and blue lines represent the electric fields $E_L$ obtained by numerically solving Eq.~(\ref{wakefield_equation1}) with chirp rates $\alpha=0$ and $\alpha=-0.0014$, respectively. The black dashed lines represent the wave-breaking limitation $E_0$ in terms of the RL limit value $E_{\rm RL}(N_e=n_e)$.

In Fig.~\ref{robustness}(a), the PIC simulation results of $E_L$ are depicted by the red diamond line with chirp rates $\alpha=-0.0014$. The broad range of $t_0$ values that allow $E_L>E_{\rm RL}$ is seen in the PIC simulations and  in the solution of Eq.~(\ref{wakefield_equation1}). The wave-breaking threshold, however, limits the amplitude of the plasma wave in the PIC simulations.

Furthermore, in Fig.~\ref{robustness}(b), the PIC simulation results of $E_L$ are shown by the red diamond line with chirp rates $\alpha=-0.0014$, and the pink square line without a chirp. When no chirp is applied, the plasma wave can only be significantly excited when the plasma densities nearly meet the matching condition, as depicted by the black solid line. The PIC simulation results represented by the pink square line confirm this prediction. However, with a frequency chirp $\alpha=-0.0014$, the excitation becomes more robust, although the wave-breaking threshold still limits the amplitude of the plasma wave in the PIC simulations. Notably, the broad density variations  $N_e\approx0.7n_e\sim1.15n_e$ can still lead to enhanced excitation with $E_L>E_{\rm RL}$. This observation is supported by both the solution of Eq.~(\ref{wakefield_equation1}) and the PIC simulation results. 

\nocite{*}
\bibliography{reference}

\end{document}